\documentclass[structabstract]{aa}

\usepackage{amsmath}
\usepackage[dvips]{graphicx}
\usepackage{txfonts}
\usepackage{float}

\begin{document}

\title{Dominance of outflowing electric currents \\ on decaparsec to
kiloparsec scales in extragalactic jets} 

\author{
Dimitris M. Christodoulou,\inst{1}
Denise C. Gabuzda,\inst{2}
Sebastian Knuettel,\inst{2}
Ioannis Contopoulos,\inst{3,4}
Demosthenes Kazanas,\inst{5}
\and
Colm P. Coughlan\inst{6}
 }

\institute{
Dept. of Mathematical Sciences,
University of Massachusetts Lowell, Lowell 01854. \\
                 E-mail: dimitris\_christodoulou@uml.edu\\
\and
Dept. of Physics, University College Cork, Cork, Ireland. \\
                 Email: d.gabuzda@ucc.ie\\
\and 
Research Center for Astronomy and Applied Mathematics,
Academy of Athens, Athens 11527, Greece. \\
                 Email: icontop@academyofathens.gr\\
\and
National Research Nuclear University, 31 Kashirskoe highway, 
Moscow 115409, Russia.\\
\and
NASA/GSFC, Code 663, Greenbelt, MD 20771. \\
                 E-mail: demos.kazanas@nasa.gov\\
\and
Dublin Institute for Advanced Studies, Astronomy and Astrophysics
Section, 31 Fitzwilliam Place, Dublin 2, Ireland.
}

\def\gsim{\mathrel{\raise.5ex\hbox{$>$}\mkern-14mu
             \lower0.6ex\hbox{$\sim$}}}

\def\lsim{\mathrel{\raise.3ex\hbox{$<$}\mkern-14mu
             \lower0.6ex\hbox{$\sim$}}}

\date {Received ; accepted }

\abstract
{Helical magnetic fields embedded in the jets of active galactic
nuclei (AGNs) are required by the broad range of
theoretical models that advocate for electromagnetic launching
of the jets. In most models, the direction of the magnetic field
is random, but if the axial field is generated by a Cosmic
Battery generated by current in the direction of rotation in the accretion disk, there is a correlation 
between the directions of the spin of the
AGN accretion disk and of the axial field, which leads to a specific 
direction for the axial electric current, azimuthal magnetic
field, and the resulting observed transverse Faraday--rotation (FR)
gradient across the jet, due to the systematic change in the
line-of-sight magnetic field.}
{We consider new observational evidence for the presence of a
 nested helical magnetic-field structure such as would be brought 
about by the
operation of the Cosmic Battery, and make predictions about the
expected behavior of transverse FR gradients observed on decaparsec 
and kiloparsec scales.}
{We have jointly considered 27 detections of transverse FR gradients
on parsec scales, four reports of reversals
in the directions of observed transverse FR gradients observed on
parsec--decaparsec scales, and five detections of transverse FR gradients 
on decaparsec--kiloparsec scales, one reported here for the first time. 
}
{The data considered indicate a predominance 
of transverse FR gradients in the clockwise direction on the
sky (i.e., net axial current flowing inward in the jet) on parsec 
scales and in the counter-clockwise direction on the sky 
(i.e., net axial current flowing outward) on scales greater than 
about 10~pc, consistent with the expectations for the Cosmic Battery.
}
{The collected results 
can be understood if the dominant azimuthal field
on parsec scales corresponds to an axial electric current flowing 
inward along the jet, whereas the (weaker) dominant azimuthal field on 
kiloparsec scales corresponds to a outward-flowing current in the outer
sheath of the jet and/or an extended disk wind. 
}


\authorrunning{Christodoulou et al. 2015}
\titlerunning{Outflowing Electric Currents in Extragalactic Jets}

\maketitle

\section{Introduction}

\subsection{Faraday Rotation in Radio Sources}

There are two ways of probing the geometry of magnetic fields in
astrophysical plasmas: (i) the intrinsic orientation of the polarization 
angle of synchrotron radiation provides information about the projection
of the synchotron magnetic field onto the plane of the sky and (ii) 
Faraday rotation (FR) --- the wavelength-dependent rotation of the intrinsic
polarization angle of linearly polarized radiation as it traverses a
magnetized plasma --- provides information about the line-of-sight
component of the magnetic field in the region of Faraday rotation.
Because Faraday rotation provides information only about the
line-of-sight magnetic field, the full three-dimensional (3D) 
structure of the field cannot be deciphered uniquely solely from 
direct or synthesized FR observations.
This drawback can, in
principle, be overcome by modeling the polarized emission regions
and the embedded magnetic field and projecting model Faraday
rotation-measure (RM)
maps onto the observer's sky to enable direct comparisons (Murgia
et al. 2004; Laing et al. 2006, 2008; Guidetti et al. 2010; Govoni
et al. 2010; Bonafede et al. 2010).  Model RM maps constructed in
this way use a minimal set of assumptions to avoid imposing
preconceived notions about the plasma and the magnetic field and
their interactions with the surrounding intergalactic environment.
This type of modeling for galaxy clusters implies mean inter-cluster 
magnetic-field
strengths of a few to several tens of $\mu$G and the presence of a
random component on scales of $\sim$1~kpc in radio galaxies that
are members of large clusters and small groups (Laing et al. 2006,
2008; Guidetti et al. 2010; Bonafede et al. 2010).

The polarization structure observed in the jets of extragalactic
radio sources can be used to construct
models for the structure of the underlying magnetic field in the jets.
Such models of the magnetic field and the FR of the
radiation emitted by the jets and the lobes of
extragalactic radio sources have been broadened in scope by including
physical assumptions about the 3D structure of the field, the
kinematics and emissivity of the jet outflow, and the presence of
cavities around the jets (Laing et al. 2006, 2008).

\subsection{Magnetic fields in jets}

It is now well understood that an active galactic
nucleus (AGN) consists of two components: a
rotating supermassive ($\sim 10^8 M_\odot$) black hole with its
event horizon extending out to $\sim$1~AU, and a surrounding
rotating accretion disk extending out to $\sim$1~pc.
It is widely believed that the physical mechanism that drives these
systems is magnetohydrodynamical: the rotating accretion disk is 
threaded by a magnetic field that
is wound up by the differential rotation of
the disk plasma (i.e., the field becomes helical), driving a collimated
outflow (jet) above and below the symmetry plane of the disk. The
asymptotic velocity attained by the jet material is of the same
order as the rotational velocity at the base of the jet. The jets
are also expected to include two components:
an inner, relativistic, axial jet and an outer, nonrelativistic,
extended disk wind (Pelletier et al. 1988; Ferreira et al. 2006).
The theoretical details of the jet launching mechanism have been
well-studied in the literature (Blandford \& Payne 1982; Contopoulos
\& Lovelace 1994; Contopoulos 1995; see also the review by
Spruit~2010 and references therein). It is expected that
the magnetic field is efficiently wound up only beyond the 
Alfv\'en distance, which is  $\sim$10 times the 
radial extent of the outflow at its base. Thus, these theoretical 
models suggest that the magnetic fields
of the inner jet and the outer extended wind will develop
significant toroidal components on distances beyond $\sim$10~AU and
$\sim$10~pc, respectively, from the base of the jet (see Fig.~1). 

\begin{figure}[t]
\centering
\includegraphics[width=0.35\textwidth,angle=90]{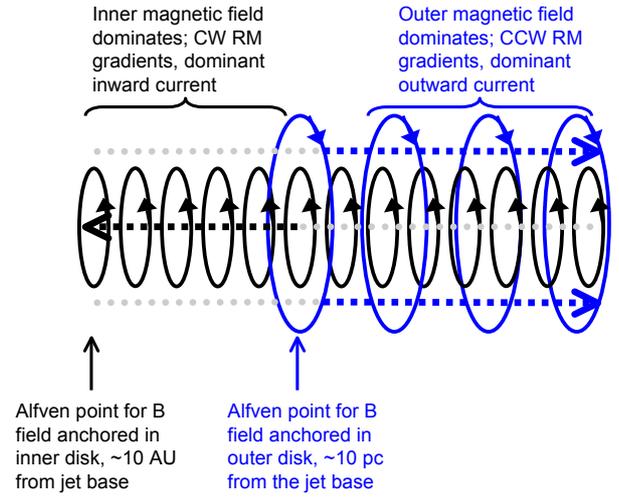}
\caption{Schematic of the key transition points along the jet axis (not to
scale): the 
Alfven point for the inner field, which lies roughly 10~AU from the jet base,
and the Alfven point for the outer field, which lies roughly 10~pc from the 
jet base. The jet base is located further to the left. 
Ellipses representing the azimuthal component of the inner field
are shown in black, and ellipses representing the azimuthal component of
the outer field in blue. In each case, the orientation of the field is 
shown by corresponding arrows. The associated currents are shown by the
dashed lines. CW RM gradients and inward currents dominate on parsec scales,
while CCW RM gradients and outward currents dominate on decaparsec--kiloparsec 
scales.} \label{Figure-Alfven}
\end{figure}

The old idea that loops of weak
field can be generated randomly by the Biermann (1950) battery in
the accretion disks, where they are amplified by turbulent dynamos,
so that their tangled fields are embedded in the outflows, has not
been successful in explaining the observations. Recent radio observations
indicate the presence of magnetic fields organized on scales of at least
$\sim$10--30 kpc in AGN (e.g., Carilli \& Taylor 2002; Widrow 2002; Eilek
2003; Kronberg 2005, 2010). Furthermore, the inefficacy of this
mechanism has been recognized in various theoretical studies (e.g.,
Vainshtein \& Rosner 1991; Subramanian 2008; Kulsrud \& Zweibel
2008).  Despite these problems, the Biermann battery continues to 
be adopted as the physical basis in investigations of magnetic-field generation in AGN, because of the perceived lack of an alternative
mechanism.

\subsection{The Cosmic Battery}

\begin{figure}[t]
\centering
\includegraphics[width=0.35\textwidth]{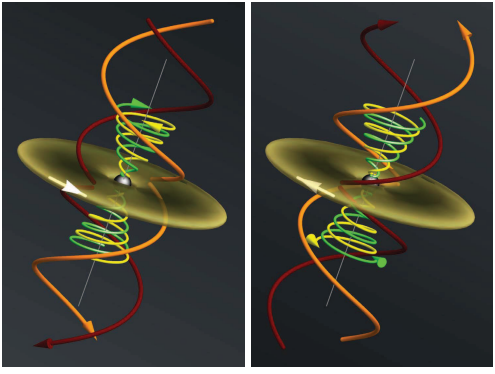}
\caption{Schematic of the helical-field structure of an AGN for
arbitrary observer orientation as predicted by the Cosmic Battery.
The central black hole is represented by the gray sphere, the surrounding accretion disk by the  flattened
yellow disk, and the axis of the system by the gray line. The white arrow indicates the direction of the
disk/black hole rotation. Inner field lines threading
the black hole are yellow and green. Outer field lines threading the
disk are orange and red. The Cosmic Battery predicts that the axial
direction of the inner/outer field is along/opposite to
$\boldsymbol\omega$, respectively (eq.~[\ref{muB}]). This corresponds
to an axial electric current flowing inward along the jet axis and
outward farther out where the outer
field lines are drawn.  Field lines are wound by the black hole/disk 
rotation. An observer will see the transverse RMs 
increasing on the sky CW/CCW relative to the jet base for the inner/outer azimuthal field.
} \label{Figure2}
\end{figure}

In the past few years, a promising alternative to the Biermann
battery has been found, which seems capable of generating strong,
ordered, large-scale magnetic fields in AGN accretion disks and of
providing the physical background needed to understand
some of the most prominent features observed in magnetized jets.
This alternative mechanism relies on the Poynting--Robertson drag
on plasma electrons to generate large--scale azimuthal currents in
the inner disks of AGN (Contopoulos \& Kazanas 1998; Contopoulos
et al. 2006; Christodoulou et al. 2008; 
Lynden-Bell~2013; Koutsantoniou \& Contopoulos~2014; Contopoulos
et al.~2015). Its most important element is that it supports and
maintains a large-scale poloidal magnetic field whose direction is
inextricably tied to the rotation of the disk (see also
Lynden-Bell~2013). In the vicinity of a $10^8 M_\odot$ black hole,
this mechanism can generate a field corresponding to equipartition
between the magnetic and relativistic-particle energies on timescales of 
the order of one billion years (Contopoulos \& Kazanas 1998; Contopoulos 
et al.~2015). We call this mechanism the Cosmic Battery (hereafter CB).

According to the CB, the Poynting-Robertson drag force on the
electrons at the inner edge of the accretion disk around an AGN's
supermassive black hole\footnote{In fact, around an accreting black
hole of any size; see e.g., Contopoulos  \& Kazanas (1998), Kylafis
et al. (2012).} generates a toroidal electric current which gives
rise to poloidal magnetic field loops around the inner edge of the
disk with a magnetic moment $\boldsymbol{\mu_B}$. The outer footpoint of
each loop resides well inside the accretion disk and diffuses
outward on the disk's local diffusion timescale, provided that this
is at least about a factor of two shorter than the local accretion
timescale. The inner footpoint is dragged inward by the accretion
flow and eventually ends up near the black hole. Under these
conditions, the magnetic flux that accumulates in the vicinity of
the black hole due to the continuous (secular) growth of the inner
field always points in the same direction as the angular
velocity vector $\boldsymbol\omega$ of the accretion disk:
\begin{equation} \boldsymbol{\mu_B} \parallel
\boldsymbol\omega\ . \label{muB}
\end{equation}
The footpoints of each poloidal loop also participate in
the rotation of the disk, thus they are both dragged along
the direction of rotation. The resulting 
wound-up field configuration corresponds to a large-scale axial
electric current ${\cal I}_{\rm inner}$ that flows toward the disk
along the symmetry axis of the inner (core) jet, i.e.,
\begin{equation}
{\cal I}_{\rm inner} \ \mbox{flows opposite to the jet direction},
\end{equation}
and a large-scale axial electric current ${\cal I}_{\rm outer}$
that outflows from the disk and into the outer wind, i.e.,
\begin{equation} {\cal I}_{\rm outer} \
\mbox{flows along the jet direction}
\end{equation}
(see Fig.~1). The central jet current closes along the interface between 
the core
jet and the outer wind, whereas the outer wind electric current
closes farther out (not along the inner core jet).
This universal configuration of the axial electric current
circuit in extragalactic jets, predicted by Contopoulos et
al.~(2009) (hereafter CCKG), is unique to the CB mechanism and it was
confirmed numerically by the simulations of Christodoulou et al.
(2008) and Contopoulos et al.~(2015).

\subsection{Observational signatures of the Cosmic Battery}

The most obvious diagnostic tool of the above magnetic-field
configuration is the observation of characteristically oriented
transverse Faraday RM gradients across the jets 
and the lobes of at least some radio sources that are not too 
disturbed by interactions with their environment or other 
internal factors. As depicted in Figs.~1 and 2, when the observed RMs
increase in the clockwise (CW) direction on the sky relative to 
the base of the jet outflow, the electric
current flows inward opposite to the jet direction, whereas if the
RMs increase in the counterclockwise (CCW) direction on
the sky relative to the jet base, the electric current flows
outward along the jet direction. In summary:
\begin{equation} \mbox{RM increases CW} \ \Longrightarrow \ {\cal I}_{\rm inner} \
\mbox{opposite to the jet direction},
\end{equation}
\begin{equation} \mbox{RM increases CCW} \ \Longrightarrow \ {\cal I}_{\rm outer} \
\mbox{along the jet direction}.
\end{equation}

These results are independent of observer location 
(see Fig.~2 and CCKG for more details). Now, as was pointed out above
(see Fig.~1),
the inner field should be efficiently wound up only outside the 
corresponding Alfv\'en point, meaning at distances greater than about 
10~AU from the jet base; in terms of observed transverse Faraday RM
gradients due to the systematic change in the line-of-sight component
of the azimuthal field, we expect the inner azimuthal field and  
${\cal I}_{\rm inner}$ to give rise to significant transverse Faraday 
RM gradients only at distances greater than about 10~AU from the jet 
base. Analogously, the outer field should be efficiently wound up only 
at distances greater than about 10~pc from the jet base; accordingly, 
we expect the outer azimuthal field and  
${\cal I}_{\rm outer}$ to begin making significant contributions to
the Faraday RM gradients only at distances greater than about 10~pc 
from the jet base. This suggests that the net Faraday rotation due 
to both the inner and outer azimuthal fields (i.e., the net transverse 
RM gradient), should be dominated by the inner azimuthal field at
distances out to about 10~pc from the jet base, then by the outer
azimuthal field starting at distances of about 10~pc or more (see Fig.~1). 
There could also be a region at distances of a few tens of pc where no
clear transverse RM gradients are observed, because the contributions
from the inner and outer azimuthal fields are comparable.

CCKG looked for precisely this effect using published RM maps
for roughly 30 AGNs whose parsec-scale jets mapped with very long
baseline interferometry (VLBI) seemed to show reasonably clear transverse 
RM gradients. CCKG reported evidence for a predominance of 
CW transverse RM gradients within $\simeq 20$~pc of the center, 
whereas CCW transverse RM gradients were found 
in some of these sources at larger distances. These results were 
disputed by Taylor \& Zavala (2010), who claimed that, in the vast
majority of cases, the VLBI jets were too poorly resolved to make
the observed transverse RM gradients reliable. However, the Monte 
Carlo simulations of Hovatta et al. (2012) and Mahmud et al. (2013) 
subsequently demonstrated that transverse RM gradients could be detected
even when the intrinsic jet width was much narrower than the beam
width for the VLBI array used, removing the doubts cast by Taylor
\& Zavala (2010). Previously firm results, recently reported new 
results and reanalyses of a number of previously published RM 
images applying the improved error estimation approach developed 
by Hovatta et al. (2012) have now brought the list of reliable 
(monotonic, with significances $> 3\sigma$) parsec-scale transverse 
RM gradients 
to 27, of which 20 are CW and 7 are CCW on the sky, relative to
their jet bases (Gomez et al. 
2008; Kharb et al. 2009; Hovatta et al. 2012; Gabuzda et al. 2014a, 
2014b, 2015a); a simple binomial probability distribution analysis 
indicates that the probability of at least 20 out of 27 of the
observed transverse RM gradients having the same orientation (CW)
by chance is about $0.95\%$,
supporting the reports of CCKG of a predominance of CW transverse
RM gradients on parsec scales. In the sizeable minority of parsec-scale
jets displaying CCW transverse RM gradients (7 out of 27), the gradients
may be present at distances from the cores that are greater 
than the transition distance of about 20~pc; a more
detailed analysis of this question will be considered by Gabuzda
et al. (in preparation). In 
addition, it is always possible that the battery mechanism
is not efficient in some jet--disk systems. 
Work is ongoing to try to add to the
list of AGN jets with reliable transverse RM gradients on parsec
scales (Gabuzda et al., in preparation).
A precise demarkation line between the distances at which the
inner and outer azimuthal magnetic fields dominate the observed
transverse RM gradients cannot be determined with certainty from 
the available
VLBI data, but the outer (return) field with its CCW RM gradients
appears to dominate at distances larger than about 20~pc (Fig.~2
in CCKG), at least in those objects in which the RM gradients are
resolved and are not obscured by substantial random RM components.

\subsection{Outline of the paper}

With the above theoretical considerations in mind, we set out to
examine the available information about transverse RM gradients on 
larger scales extending out to kiloparsecs, exceeding the expected
distance of the Alfv\'en point for the outer azimuthal magnetic
field from the jet base. The purpose of
this study is essentially twofold: (i) to establish whether or not
transverse RM gradients reasonably interpreted as reflecting the
presence of helical jet magnetic fields are present on scales of
tens to thousands of parsec and, if present, (ii) to establish 
whether such RM
gradients show any evidence for a preferred direction. 

In Section~2, we describe observations for three sources we
have analyzed for this paper.
In Section~3, we consider three previous firm reports of transverse 
RM gradients on scales exceeding about 20~parsec, for which quantative
analyses have already been carried out in previous publications, 
to which we add
our new detection of a $3.0\sigma$ transverse RM gradient in A2142A. We
also consider
four reliable cases of reversals in the direction of the observed RM 
gradients with distance from the jet base. Finally, we consider an
additional seven transverse RM gradients that are visible in previously published 
RM maps, and present the results of quantitative analyses of the RM results
for three of these sources. All seven of these transverse RM gradients
must be considered tentative, due to insufficiently high significance
of the gradients ($< 3\sigma$) , lack of information about the uncertainties 
in the observed RM values, and/or uncertainty in the orientation of the
gradients relative to the local jet direction.
In Section~4, we discuss the implications of these collected
results both in general and in the context of the CB mechanism.
Finally, we conclude in Section~5 with a
summary of our results and some remarks about related ongoing
research of the large-scale magnetic fields in AGNs and in our own
Galaxy.

\section{Observations and reduction}

\subsection{VLA Data}

A. Bonafede and F. Govoni kindly provided the calibrated visibility data 
that had been used to make the published RM maps for A2142A (Govoni et al. 
2010) and 5C4.152 (Bonafede et al. 2010).  Data are available for 5C4.152 
at 4.635, 4.835 and 8.275~GHz, and for A2142A at 4.535, 4.835, 8.085, and 
8.465~GHz. 
The observations and data calibration and reduction methods used 
in the initial analyses carried out for these objects are given by 
Bonafede et al. (2010) and Govoni et al. (2010).

We could not use the RM maps published by Bonafede et al. (2010) and
Govoni et al. (2010) directly, because the associated error maps did not 
take into account the finding of Hovatta et al (2012) that the uncertainties 
in the Stokes $Q$ and $U$ fluxes in individual pixels on-source are 
somewhat higher than the off-source rms fluctuations, potentially 
increasing the resulting RM uncertainties.

To address this, we imported the final, fully self-calibrated
visibility data into the {\sc AIPS} package, then
used these data to make naturally weighted $I$, $Q$ and $U$ maps
at all wavelengths, with matching image sizes, cell sizes and beam
parameters specified by hand in the {\sc AIPS} task {\sc IMAGR}.
These images were all convolved with a circular Gaussian beam having a
full-width at half-maximum of 3$^{\prime\prime}$. We obtained
maps of the polarization angle, $\chi = \frac{1}{2}\arctan(U/Q)$, and
used these to construct corresponding RM maps in the {\sc AIPS} and
{\sc CASA} packages.  The uncertainties in the polarization
angles used to obtain the RM fits were calculated from the uncertainties in $Q$
and $U$, which were estimated using the approach of Hovatta et al. (2012).
In all cases, satisfactory RM fits were obtained without applying
$n\pi$ rotations of the observed polarization angles.

\subsection{VLBA data}

We also analyzed the RM distribution of 3C120, constructed using data 
obtained with the Very Long Baseline Array (VLBA) at 1.358, 1.430, 1.493 and 
1.665~GHz.
This RM map was originally published by Coughlan et al. (2010); again, we
were not able to use the previously published RM map directly because
the associated error maps did not take into account the improved method
of Hovatta et al. (2012) for estimating the on-source $Q$ and $U$ 
uncertainties in individual pixels.

We addressed this by using the $Q$ and $U$ maps used by Coughlan et al.
(2010) to produce new RM maps in both {\sc AIPS} and {\sc CASA}, assigning
$Q$ and $U$ uncertainties in accordance with the approach of Hovatta et 
al. (2012).

In all cases, when differences in RM values across the jet were obtained, the 
uncertainties of the RM values did not include the effect of uncertainty 
in the polarization angle calibration, since this cannot introduce spurious 
RM gradients (Mahmud et al. 2009, Hovatta et al. 2012).  The
uncertainty of the difference between the RM values at the two ends of
a slice was estimated by adding the uncertainties for the two RM values
in quadrature.  

\begin{figure}
\begin{center}
\hspace*{-1.0cm}
\includegraphics[width=.45\textwidth,angle=0]{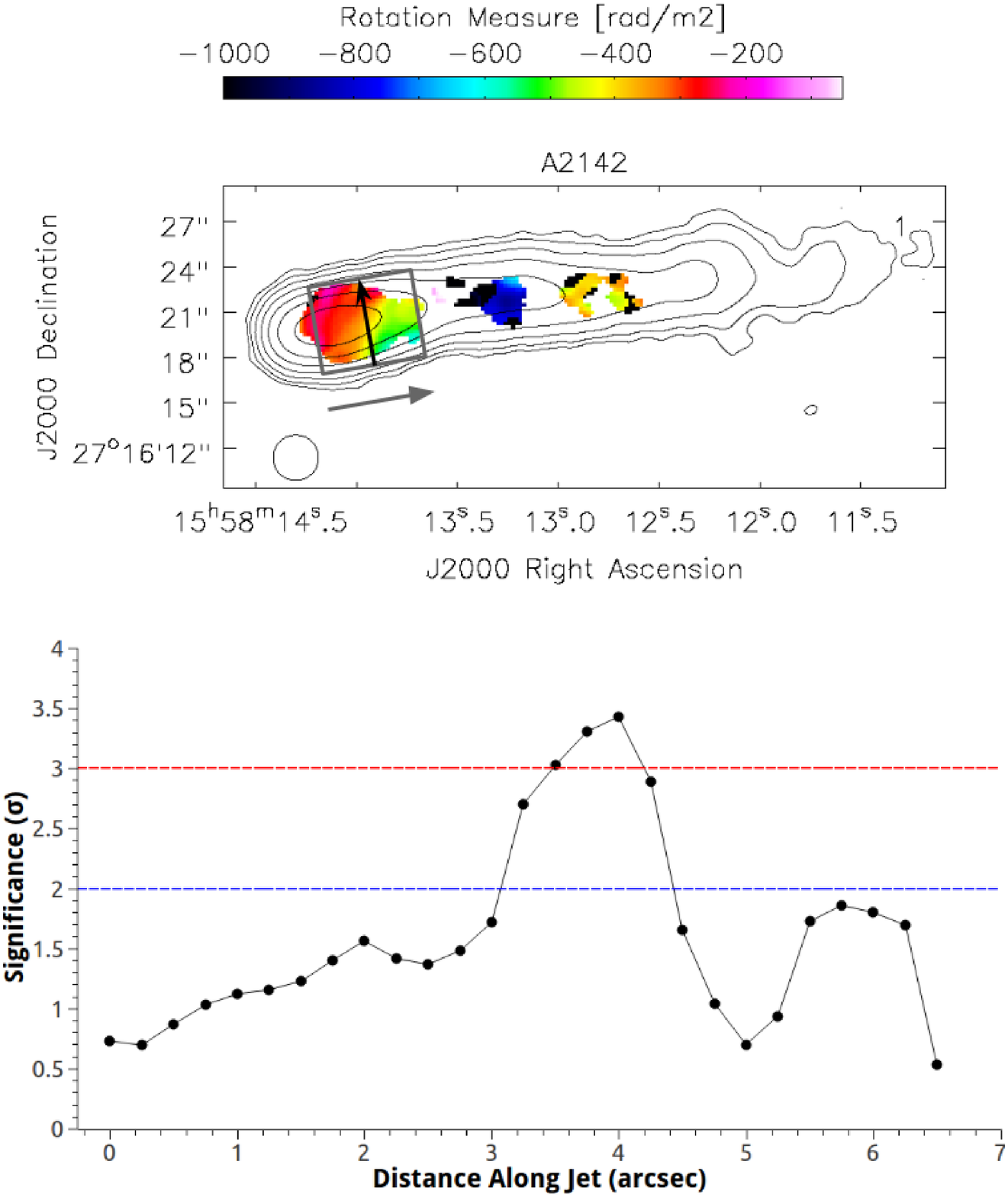}
\end{center}
\hspace*{0.3cm}
\caption{Intensity map at 4.535~GHz  of A2142A with 
the RM distribution
superposed (upper panel). The lowest contour is 1\% of the peak intensity of 8.8~mJy/beam, 
the contours increase in increments of a factor of two, and the beam size
is shown in the lower left corner of the image.  The black arrow across 
the jet highlights the direction of the transverse RM gradients.  The 
gray box shows the region for which the significances of series of 
parallel, monotonic transverse RM gradients are plotted in the lower 
panel; the gray arrow outside the box pointing outward along the jet 
shows the direction 
of increasing pixel number in the lower panel, and pixel~0 corresponds 
to the inner edge of the gray box.  Pixel size is 0.25~arcsec.  
The $2\sigma$ level is shown by the dashed blue horizontal line, and the 
$3\sigma$ level by the dashed red horizontal line.}
\end{figure}

\section{Transverse RM gradients on scales exceeding $\simeq$~20~pc}

\subsection{Previously published firm gradients}

Virtually all the work done on transverse RM gradients has been
carried out for high-resolution data obtained on the VLBA at wavelengths between 15 and 5~GHz; most of the 
RM gradients detected cross their jets at projected distances from the 
jet base of no more than a few milliarcseconds (less than about 
20--30~pc). 
Very few studies probing larger scales have been carried out, and
we will now consider the few results available here. In all cases,
the statistical significances of the transverse RM gradients have
been reliably estimated, and found to be at least $3\sigma$.
When we refer to a transverse RM gradient being
CW or CCW, we mean that its orientation is clockwise or 
counter-clockwise on the sky, relative to its own jet base.

\begin{enumerate}
\item {\em 1652+398 (Mrk~501)}.---The RM map of this AGN based on VLBA data at 
frequencies between 1.6 and 8.4~GHz published by
Croke et al. (2010) shows a very clear transverse RM gradient 
throughout the extended jet before it turns sharply about 30~mas  
(about 20~pc projected distance) from the core, oriented CCW. Croke et al.
(2010) present a quantative analysis of this gradient. 
Even taking into account the fact that the RM uncertainties indicated 
by Croke et al. (2010) were not obtained using the improved method
of Hovatta et al. (2012), and therefore could be up to about a factor 
of two too small, the significance of this gradient is far in excess 
of $3\sigma$.

\smallskip
\item {\em 3C380}.---The RM map of this AGN based on VLBA data at wavelengths 
between 1.4 and 5.0~GHz published by Gabuzda et al. (2014a) shows a 
transverse RM gradient oriented CCW; the quantative analysis carried
out by Gabuzda et al. (2014a) shows that this gradient has a significance 
of about $4\sigma$. 

\smallskip
\item {\em 5C4.114}.---An arcsecond-scale VLA RM map for the kiloparsec-scale 
jets of 5C4.114 recently analyzed by Gabuzda et al. (2015b) demonstrates 
transverse RM gradients across both the northern and
southern jets with significances of about $4\sigma$ and $3\sigma$,
respectively, both oriented CCW. 
\end{enumerate}

\subsection{New detection of a firm transverse RM gradient in A2142A}

The radio source A2142A is located in a cluster, and we therefore
expect some contribution to the RM distribution from the magnetized
intercluster gas; this should most likely not show any large-scale order,
and be predominantly ``patchy''.  In addition to some patchiness, the 
RM distribution of A2142A shows a 
tendency for the RM values along the northern side of the jet to be
less negative than those along the southern side (Fig.~13 of 
Govoni et al. 2010), corresponding to a possible gradient in the RM
values across the jet. We note that the core is at the eastern end of 
the observed radio structure, so that this implied transverse RM
gradient is oriented CCW. 

Fig. 3 presents our 4.535-GHz intensity image of A2142A, with the RM image
superposed in color; this essentially reproduces the images in Fig.~13 
of Govoni et al. (2010). The output pixels in the RM map were blanked when the 
RM uncertainty resulting from the $\chi$ vs. $\lambda^2$ fits exceeded 
80~rad/m$^2$. Our analysis of the entire RM distribution showed the 
presence of monotonic RM gradients across the jet in the region surrounded 
by the gray box. At the
redshift of A2142A, $z =  0.091$, these correspond to projected distances of
roughly 10~kpc from the jet base. The points in the lower panel 
of Fig.~3 correspond to monotonic transverse RM gradients obtained 
for a series of parallel RM slices across the jet, inside the gray box in
the upper panel. 
As can be seen, comparisons 
of the RM values at the two ends of the RM slices considered indicate that 
the transverse RM gradients about 4~arcsec from the start of the boxed region
have significances reaching about $3.5\sigma$, with another region of 
gradients reaching nearly $2\sigma$ slightly further out from the core.
Thus, we consider this a firm case of a transverse RM gradient on 
kiloparsec scales.

\subsection{Previously published firm RM-gradient reversals}

To the four sources in Sections~3.1 and 3.2, we add information from reports of
reversals in the directions of the observed transverse RM gradients
in four more AGNs: 0716+714 (Mahmud et al. 2013, reversal at a projected
distance of a few pc from the jet base), 0923+392 (Gabuzda et al. 2014b, 
reversal at a projected distance of about 15~pc from the jet base), 
1749+701 (Mahmud et al 2013, reversal at a projected distance of about 
35~pc from the jet base), and 2037+511 (Gabuzda et al. 2014b, reversal about
35~pc from the jet base). Although the observed reversal in the 
direction
of the observed transverse RM gradient in 0716+714 was relatively close
to the core, the 1.4--1.7~GHz data analyzed by Healy (2013) show that the
orientation of the outer gradient was maintained to projected distances 
of about 35~pc from the core. In all four cases, quantitative analyses
carried out in the papers cited above indicate that both transverse RM
gradients detected had significances of at least $3\sigma$, and
the inner gradient was oriented CW and the outer gradient CCW relative
to the jet base.

\subsection{Tentative gradients from an initial inspection of RM maps 
in the literature}

The transverse RM gradients in 5C4.114 and A2142A listed in Sections~3.1
and 3.2 were initially identified via an initial inspection of 
published Faraday RM maps 
of extragalactic radio sources on kiloparsec scales, carried out for
85 objects 
(Athreya et al. 1998; Best et al. 1998, 1999; Feretti et al.
1999; Venturi \& Taylor 1999; Taylor et al. 2001; Eilek \& Owen
2002; Goodlet et al. 2004; Govoni et al. 2006; Laing et al. 2006,
2008; Kharb et al. 2008; Guidetti et al. 2008, 2010; Feain et al.
2009; Kronberg 2009; Bonafede et al. 2010; Govoni et al. 2010;
Kronberg et al. 2011; Algaba et al. 2013). 
In all cases, the authors of these
studies took care to ensure that the RM images were constructed
using only data with sufficiently high signal-to-noise ratios, and
that the resulting $\lambda^2$ fits for the RM values were
sufficiently good.  Unfortunately, full information about the 
uncertainties in the RM values at individual points in the RM
maps is not available; therefore we were able to identify
candidate transverse RM gradients, but have not been able to carry out
quantitative analyses to test their significances. 

We were unable to identify any obvious large-scale monotonic
transverse RM gradients in 77 of these 85 published
maps, in which (a)~the jets appeared to be strongly influenced by
their surroundings (they were strongly bent or appeared to be
partially disrupted); (b)~the RM distribution was patchy; (c)~limited to
just a few pixels across or along the images (insufficient resolution);
(d)~the RM images were based on only two
frequencies; or (e)~the results were shown in gray scale that did
not make the structures in the RM images sufficiently clear. 
We sought to obtain higher-resolution or
color images from some authors, and we appreciate their efforts to
provide us with the maps we requested. 

In the end, we identified eight potential objects whose RM maps showed
extended, monotonic, nearly transverse RM gradients. Of these, 
two (5C4.114 and A2142A) have now been shown to be
statistically significant (Gabuzda et al. 2015b; this paper), leaving 
six additional tentative gradients. To these we add results for 3C120 
on slightly smaller scales of 30--100~pc presented by Coughlan 
et al. (2010). We have not attempted to measure or analyse the
magnitudes of these tentative RM gradients, which would be
extremely difficult to interpret unambiguously due to convolution with the 
observing beam. Instead, we focus on the presence of transverse gradients 
visible in the RM distributions and their directions relative to 
the bases of the jets, quantatively estimating the significances 
of these gradients when possible.

We briefly describe the RM distributions of these seven sources, and
the basis for our suggestion that these may contain transverse RM
gradients.  The patchiness of many of the 85 published RM distributions we considered is consistent with a picture in which random distributions 
of the magnetic field and electron density in the general vicinity of the 
radio source (e.g., in the cluster or inter-cluster medium in which the 
source is located) generally dominate on these large scales; these seven 
sources appear to be those in which an ordered RM pattern is dominant
in at least some regions.
As above, when we refer to a transverse RM gradient being
CW or CCW, we mean that its orientation is clockwise or 
counter-clockwise on the sky, relative to its own jet base.

\begin{enumerate}
\item {\it 0156$-$252}.---There appears to be an RM gradient
across the eastern jet (Fig.~3 of Athreya et al.~1998), oriented in the CCW
direction. Unfortunately, we do not have access to these data and were
not able to quantatively analyze the significance of this gradient. 
Therefore, this remains a tentative transverse RM gradient.

\smallskip
\item {\it 3C120}.---In contrast to the other objects considered
in this section, the Faraday-rotation map of Coughlan et al.
(2010) is based on VLBA data at four frequencies between 1.4 and
1.7~GHz, and thus probes slightly smaller angular scales. Transverse RM 
gradients are present at several locations on scales of approximately 
30--100~pc from the core, all oriented CCW. 

\begin{figure}[H]
\begin{center}
\includegraphics[width=.45\textwidth,angle=0]{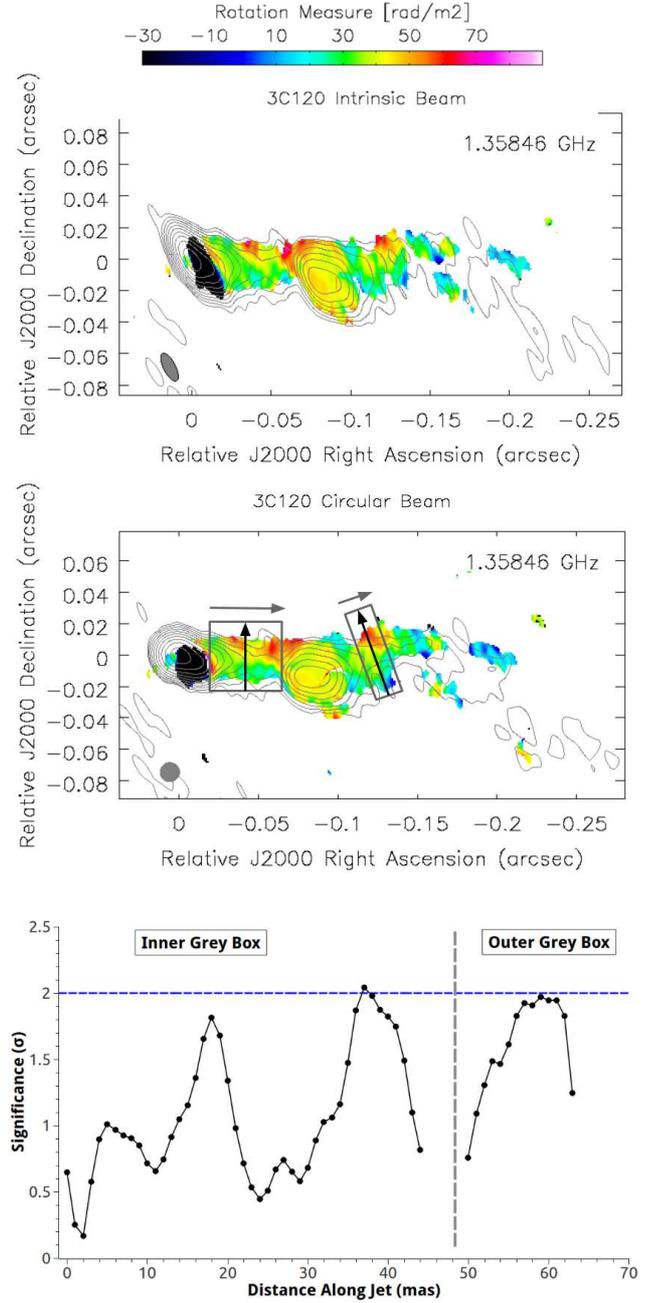}
\end{center}
\hspace*{0.3cm}
\caption{Top panel: Intensity map at 1.358~GHz  of 3C120 convolved
with the ``intrinsic'' beam and the RM distribution superposed. The lowest 
contour is 0.125\% of the peak of 1.37~Jy/beam.  The middle panel
shows the same intensity and RM maps convolved with a circular beam of equal
area; the lowest contour is 0.125\% of the peak of 1.43~Jy/beam. In both cases,
the contour step is a factor of two, the pixel size is 0.25~arcsec, and
the beam size is shown in the lower left corner of the image.  
The bottom panel shows a plot of the significances of a series of transverse
RM gradients calculated inside the gray boxes in the middle panel; the 
$2\sigma$ level is shown by the dashed blue horizontal line. See text for 
more detail.}
\end{figure}

\smallskip
The top panel of Fig. 4 presents our 1.358~GHz intensity and RM images of 
3C120 (beam size $19.6~\textrm{mas} 
\times 7.6~\textrm{mas}$ in position angle $27^{\circ}$); these essentially 
reproduce the images of Coughlan et al. (2010). The middle panel gives 
a version of the same RM map convolved with a circular beam (beam 
radius 12.2~mas). 
In both cases, the output pixels in the RM maps were blanked when the 
RM uncertainty resulting from the $\chi$ vs. $\lambda^2$ fits exceeded 
20~rad/m$^2$.  Both maps show very similar RM structures, and we 
present our further analysis for the map convolved with the circular beam.

\smallskip
Our analysis of the entire RM distribution showed the presence of 
monotonic RM gradients across the jet in the regions surrounded
by the gray boxes. 
The black arrows in the middle panel highlight the main regions of 
transverse RM gradients.  
The points in the bottom
panel of Fig.~4 correspond to monotonic transverse RM gradients obtained
for a series of parallel RM slices across the jet, inside the gray boxes in
the middle panel.
The gray arrows pointing outward along the jet in the middle panel
show the direction of increasing pixel number in the lower panel: pixel~0 
corresponds to the inner edge of the left-hand gray box, and pixel~50 to the
inner edge of the right-hand gray box.  
We can see that these slices reach significances $\simeq 2\sigma$ at
several locations along the jet, but none reach $3\sigma$.  Thus, 
these transverse RM gradients cannot yet be considered firm. However, 3C120
remains on the list of sources displaying tentative transverse RM 
gradients; deeper VLBA observations with a wider frequency range could 
help determine whether these gradients are statistically significant.

\smallskip
\item {\it M87}.---Algaba et al.~(2013) suggest that there are clear
RM gradients across HST-1 and knots A and C. The gradients are all
directed CCW.  Until a more detailed quantitative analysis is 
available, this remains a tentative transverse RM gradient.

\smallskip
\item {\it 5C4.152}.---This source has an intriguing RM
distribution, with RM gradients present across both lobes
(Fig.~15 of Bonafede et al.~2010). The gradient across the
southern lobe seems to be fairly orthogonal to the jet direction,
while the gradient across the northern lobe appears to be offset
from orthogonality as the jet bends toward the east just before
the hot spot. If the gradient across the southern lobe can be taken
to be transverse to the jet, its direction is CCW. 

\smallskip
We reconstructed the previously published RM map in order to estimate
the significance of these RM gradients.  Our quantative analysis 
of a series of RM 
slices taken across the southern lobe indicates that the strongest
transverse RM gradients have significances of about $1.5\sigma$. 
This demonstrates that these gradients cannot be considered 
statistically significant. However, we suggest that it is appropriate to 
retain 5C4.152 on the list of sources displaying tentative transverse RM 
gradients, in particular, because the 8-GHz data of Bonafede et al. (2010)
had an atypically high noise level, and the uncertainties in the RM 
values were relatively high as a result of the limited range and number of frequencies 
used (8.2~GHz, 4.6~GHz, 4.8~GHz).  Analysis of
kiloparsec-scale observations with lower noise levels and/or a wider 
frequency range could help determine more conclusively whether these 
gradients are significant or not.

\smallskip
\item {\it Cen A}.---After subtracting the overall mean
RM (which is presumably foreground), the residual RM map shows a
tendency for positive residuals on the eastern side and
negative residuals on the western side of the entire northern
lobe (Feain et al.~2009). This seems to imply an RM gradient that
extends for some 235~kpc across the northern lobe (Fig.~7 of Feain 
et al.~2009), oriented CCW. The southern lobe is strongly 
bent and no systematic trends are visible. 

\smallskip
Although there appears to be a difference in the dominant signs
of the residual RM values on either side of the northern jet axis, the
RM values themselves do not form a clear gradient across the jet.
In addition, RM measurements are available only at the positions
of background sources, so that we do not have measurements of a continuous
RM distribution across the source. Nevertheless, we  were able to test 
whether the difference in the dominant RM signs was statistically significant 
by dividing the northern jet/lobe into three sections across the jet,
encompassing the declination range  from $-38^{\circ}$ to $-41^{\circ}$
and the right-ascension ranges 13$^h$\,15$^m$0$^s$--13$^h$\,21$^m$59$^s$
(western side), 13$^h$\,22$^m$0$^s$--13$^h$\,29$^m$59$^s$ (central region), 
and 13$^h$\,30$^m$0$^s$--13$^h$\,36$^m$59$^s$ (eastern side). We then
determined the number of positive and negative RM values in the eastern
and western sections; this yielded 15/21 negative RM values in the
eastern section and 15/21 positive RM values in the western section. 

\smallskip
A simple binomial probability analysis indicates that the probability of
obtaining 15 or more of 21 values of a particular single sign is
3.92\%; if we are interested in the probability of having this fraction
of either sign, this must be multiplied by two, and so increases to 7.84\%. 
However, the probability of obtaining 15 or more of 21 values of one sign 
on one side of the jet, and simultaneously 15 or more of 21 values of 
the opposite sign on the other side of the jet, is the product of these 
the two separate probabilities, or $(0.0784)(0.0392)(100) = 0.31\%$,
which corresponds to $3\sigma$.

\smallskip
Thus, our quantative analysis demonstrates that the asymmetry in 
the signs of the RM values across the northern jet of Cen~A is statistically 
significant at the $3\sigma$ level.  However, because it is not possible to 
test whether this transverse trend in the observed RM values is monotonic 
and systematic due to the availability of RM measurements only at the
positions of background sources, we 
retain Cen~A on the list of sources whose jets display tentative
transverse RM gradients.

\smallskip
\item {\it 3C303}.---Kronberg et al.~(2011) report the presence of a
transverse RM gradient at the location of knot~E3, and use information 
about this gradient together with certain assumptions to estimate the 
associated current in the jet. The RM gradient is clearly visible 
in Fig.~3 of Kronberg et al. (2011), although it is quite narrow
and is not well resolved. The direction of the gradient is CCW. Although
this RM map has been published in the refereed literature, we
include 3C303 in our list of tentative gradients because the original
paper does not provide information about the uncertainties of the 
single-pixel RM values that can be used to estimate the significance 
of the RM gradient. We do not have access to these data,
and so were not able to carry out our own quantitative analysis 
for this source.

\smallskip
\item {\it 3C465}.---The southern jet provides a clear
example of a systematic, extended transverse RM gradient that
extends along most of this jet (Fig.~6 of Eilek \& Owen~2002).
The color scale chosen for the RM distribution is such
that the RM map presented by Eilek \& Owen (2002) is essentially
an RM sign map. The orientation of the RM gradient across the 
southern jet is CCW. The northern jet shows signs that it
intrinsically had an overall pattern, but the jet and the RM
pattern have become distorted. We do not have access to these data,
and so were not able to carry out our own quantitative analysis 
for this source.
\end{enumerate}

Table~1 summarizes information about all the firm (upper rows)
and tentative (lower rows) transverse RM gradients considered in
this Section. The references in bold in Table~1 indicate the
papers in which quantitative analyses are carried out to determine
the statistical significances of these gradients.
Treating the detections of the transverse RM gradients across the northern
and southern jets of 5C4.114 as independent measurements, we have
in total nine firm cases of transverse RM gradients. 
All these transverse RM gradients are oriented CCW 
relative to their jet bases on comparatively
large scales, greater than about 20~pc. 

\section{Discussion}

\subsection{Transverse RM gradients on scales out
to kiloparsecs from the central AGN}

The most fundamental implication of the results discussed above 
is that transverse Faraday rotation gradients --- predicted to 
exist if the jets carry toroidal or helical magnetic fields and 
detected earlier in some 27 AGNs on parsec scales --- are
also present on considerably larger scales, extending out to 
thousands of parsecs (e.g., 5C4.114; Gabuzda et al. 2015b). This 
is an important result, 
because if AGN jets carry helical magnetic fields, these 
should be present on essentially all scales where the jets propagate, 
provided that the intrinsic field structure of the jet is not disrupted 
by interactions with the surrounding environment. We note that this
conclusion follows from the firm results listed in Sections~3.1--3.3, 
and does not depend on the tentative identifications of
transverse RM gradients considered in Section~3.4.

At the same time, our literature search for transverse RM
gradients on kiloparsec scales yielded only eight tentative cases out of 85 objects considered, indicating that
it is comparatively difficult to detect transverse RM gradients 
due to the presence of helical magnetic
fields in kiloparsec-scale jets. In fact, it is easy to understand
why this should be the case: there is an
appreciable turbulent, inhomogeneous component to the thermal ambient
media surrounding the jets on these scales, which superposes a more or
less random pattern over the systematic pattern due to the helical fields.
This random component in the RM distribution apparently
dominates in the majority of
cases. This makes it perfectly natural that most of the observed RM distributions
appear random and patchy, but the overall pattern due to the helical fields
sometimes comes through. This suggests that, on average, it may be easier
to detect the systematic RM component due to helical jet magnetic fields on
parsec scales, where the ordered inner field is more dominant, a result that
seems to be borne out from the observations (see also Section~4 below).

The tentative transverse RM gradients that we consider in
Section~3.4 were not, in most cases,
noticed or appreciated by the authors of the original papers; this is 
primarily due to the fact
that their main interest was in the magnetic fields of the clusters in which the radio
sources are located, rather than in the individual radio sources themselves.
Eilek \& Owen (2002) did note the striking RM distribution in 3C465, but they
did not consider the possibility that it is associated with an embedded magnetic field
because the observed Faraday rotation is external to the main radiation source.
Bonafede et al. (2010) similarly suggested that the observed RM distributions are
due primarily to Faraday rotating material that is not associated with the radio
sources themselves, based in part on the observation that the observed Faraday
rotation is external; however, their arguments are based on general
considerations and statistical relations, rather than on an individual examination
of particular objects. Govoni et al. (2010) essentially assumed that the observed
RMs are associated only with the intracluster medium, and used the RM observations
to deduce the properties of this medium.

In our interpretation, the large-scale Faraday rotation for
the vast majority of extragalactic radio sources is indeed
dominated by turbulence and fluctuations in the cluster media; but
the eight sources  we have identified as displaying evidence for
monotonic, extended, transverse RM gradients on scales exceeding about
20~parsec (including 5C4.114; 
Gabuzda et al. 2014b) essentially represent
a handful of objects in which the dominant contribution to the
Faraday rotation in some regions is due to material associated
with the radio source itself, rather than the patchy cluster
media. In these cases, the observed Faraday rotation can still be
external (not occurring throughout the radiating volume of the
source), but it is nevertheless associated with regions carrying
the imprint of a helical magnetic field in the immediate vicinity 
of each AGN jet. Direct evidence that the observed
Faraday rotation is external to the main jet volume is indicated by
the fact that the RM fits that were obtained when constructing the RM 
maps do not show significant deviations from $\lambda^2$ behavior,
within the uncertainties (see, e.g., the plots presented by Bonafede et 
al (2010), Croke et al. (2010), Govoni et al. (2010), Mahmud et al.
(2013), and Gabuzda et al. (2014a)). 
In some cases, the three-dimensional structure of
the emitting regions may not be entirely clear; for example, it may
not be obvious whether the observed regions of the Faraday-rotation 
gradients are associated with the outflowing jet or a possibly 
back-flowing lobe structure. However, in either case, these 
regions could carry the imprint of the helical magnetic field 
carried by the jet, so that this uncertainty does not affect the
basic interpretation of the Faraday-rotation gradients we offer
here. 

\subsection{Predominance of CCW RM gradients on large scales}

A striking feature of the results considered in Section~3.1--3.3
is that there is an obvious predominance of CCW RM gradients: 
all nine firm gradients on scales exceeding about 20~pc, for all of
which quantative analyses have been carried out, are CCW 
(upper part of Table~1). Based on a simple unweighted binomial probability 
function, the probability for all nine of these gradients to be 
CCW (in agreement with the CB mechanism) by chance is about 0.2\%, 
which corresponds to about $3.1\sigma$. 

In addition, all seven tentative gradients (Section~3.4, lower part 
of Table~1) are CCW. Although the statistical significances of
most of these gradients are not known as we could not carry out 
quantitative analyses of these data, this again appears not to be random, 
in the same sense. 
The probability that all seven of these tentative gradients
would be CCW by chance is about 0.8\%, corresponding to about
$2.65\sigma$.  This suggests that at least some of these 
currently tentative RM gradients may well eventually be proven to be 
statistically significant, further strengthening the evidence for 
the presence of a helical or
toroidal field component in these AGN jets on kiloparsec scales. We note
that one factor limiting the uncertainties in the RM measurements is
the requirement that we work with the RM values in individual pixels.
In principle, it should be possible to decrease the RM uncertainties by
averaging over some number of neighboring pixels, on a scale that
remains significantly smaller than the beam size; however, this is 
not possible in practice as we do not fully understand 
the correlations between the uncertainties in neighboring pixels. Some
initial progress on this problem has been made (Coughlan 2014), but 
more work remains before it will be possible to carry out such averaging 
while also obtaining accurate estimates of the uncertainty on the average.

Thus, the results for the nine firm transverse RM gradients are 
statistically significant at the $3\sigma$ level, and this significance 
will increase if any of the tentative transverse RM gradients 
we have identified are subsequently shown to be firm detections.  
In all four cases when reversals of the RM gradients 
are observed along the jet of a single AGN (sources in Table~1 marked 
with an asterisk), the gradients are CW on small scales and CCW on larger 
scales, consistent with the prediction of the CB mechanism. 
The obviously non-random distribution of
the observed transverse RM gradients' orientations relative to
their jet bases leads us to take seriously the idea that there 
is a preferred orientation of transverse
RM gradients associated with helical jet magnetic fields.

Although this idea may seem
strange at first, it has a very natural interpretation, as we have already discussed in Section~1.4. The direction
of the RM gradient implies a direction for the azimuthal magnetic-field
component giving rise to the RM gradient. This azimuthal field 
component, in turn, implies a certain direction for the dominant current
flowing in the jet --- either inward or outward. Thus, the evidence
we find for a preference for CCW transverse RM gradients on scales
greater than about 20~parsec implies a
preference for the dominant currents in the large-scale jets of AGN to 
flow outwards from the AGN centers. In fact, this is predicted by
the Cosmic Battery mechanism (Eqs.~(3) and~(5) above), since the regions of 
the transverse Faraday-rotation gradients in most of the
sources considered in Sections~3.1--3.3 lie beyond the expected
Alfv\'en points for the inner region corresponding to the magnetized jet
outflows.

The preponderance of CCW transverse RM gradients and the presence
of helical magnetic fields on large scales can be explained
physically in the framework of the CB outlined in Section~1 above: the
return helical field dominates the total observed Faraday rotation
on relatively large angular scales on the sky, where the detected
radio emission extends to fairly large distances from the jet
axis. It is much less likely for such observations to be dominated
by Faraday rotation due to the inner helical field whose field
lines are clinging close to the jet axis. In contrast, as was shown by
the analysis of CCKG, the inner helical field is more likely
to dominate the RM measurements on parsec scales.

In the above picture, one would still not expect to
find only CW transverse RM gradients on parsec scales, i.e., inside the
Alfv\'en point, and only CCW transverse RM gradients on
kiloparsec scales, beyond the Alfv\'en point, since
various physical and observational factors perturbing the jet and
its magnetic field are bound to play a role (e.g., Broderick \&
McKinney~2010). This may explain why a minority of the sources studied
on parsec scales have displayed CCW transverse RM gradients. The exact 
distance from the jet base where the
transition from CW to CCW transverse RM gradients occurs may also
vary from source to source, so that a clear-cut demarkation line may
not exist.

Finally, CCW RM gradients are also expected when they are observed
right on top of termination shocks, provided that the shock fronts
have not been bent too much away from orthogonality to the
direction of jet propagation. Such localized CCW RM gradients may
be present in the hot spots of the FR II sources 5C4.74 and
5C4.152 (Bonafede et al. 2010), although the statistical significance
of these gradients remains to be tested.

\subsection{A theoretical alternative}

K\"onigl (2010) has argued that a predominance of CW RM gradients
on parsec scales could be the result of ordered, large-scale,
magnetic fields that were produced in the outer weakly ionized
accretion disks by Hall currents, and that were launched from the
disks in centrifugally driven wind outflows. This model may be
able to give rise to a predominance of CCW RM gradients on larger
scales, but only if the overall observed Faraday rotation is
dominated by the contribution of a ``return field'' whose origin
is not clear. Furthermore, Hall currents are usually believed to be
unimportant in AGN physics, because, unlike protostellar disks
(Krasnopolsky et al. 2011), the accretion disks of AGN (especially
those associated with the VLBA sources) are thought to be highly
ionized, in particular near the compact object where all the
prominent magnetohydrodynamical phenomena associated with the 
parsec-scale emission are thought to originate (e.g., Gaskell 2009, 
2010). This view is supported by the highly ionized oxygen ions observed 
in certain broad-line radio-galaxies, i.e., radio loud AGN viewed at
fairly small angles to the jet direction (e.g., K\"onigl et al.
1995). On the other hand, if the magnetic field is brought in to
the outer, cold, weakly ionized part of the disk from farther out
(K\"onigl 2010), then different polarities are likely to undergo
fast reconnection long before a strong toroidal field component
can develop because of the much slower differential rotation of the
outer disk.  Therefore, we consider this model physically less
plausible than the Cosmic Battery for AGNs.

\section{Summary and concluding remarks}

We have considered nine firm (having significances of at least $3\sigma$) 
and seven tentative (visible in the RM maps but whose significances are 
either below $3\sigma$ or unknown)
cases of monotonic transverse RM gradients detected across the jet 
outflows of AGNs and radio galaxies on scales exceeding about 20~pc, 
listed in Table~1. These observations provide 
direct evidence that these jets carry helical magnetic fields, whose
toroidal component sometimes survives to distances of hundreds or even 
thousands of parsec from the central AGN. 

On both the relatively large scales considered here and on parsec 
scales, transverse RM gradients have been reported for only some
fraction of the observed sources.
The simple reason why transverse RM
gradients should in fact not always be observed, even if all
AGN jets carry helical fields, is that the systematic pattern in the
RM distribution due to the helical magnetic field can be disrupted
by interactions with and entrainment of the ambient medium
through which the jet is propagating, patchiness of this surrounding
medium, and turbulence in the outer layers of the jet. The higher
detection rate of transverse RM gradient on parsec scales may then
indicate that these effects are less important and less disruptive
on smaller scales.

The data presented in Table~1 show a preponderance of large-scale transverse
RM gradients in which the RMs increase CCW on the sky relative to the jet bases,
corresponding to a dominant outward current along the associated jet structures.  
The significance of this result is currently just above $3\sigma$, and
this tendency is clear enough to warrant further study, both observational and
theoretical.

Together with the results for smaller (parsec, VLBA) scales cited in 
Section~1.4, the collected results considered in this paper are consistent with
the prediction of the CB model that CCW transverse RM gradients (corresponding 
to outward currents) should be visible across AGN jets on relatively large
scales, while CW transverse RM gradients (corresponding to inward currents) should 
be found closer to the centers of activity. The analysis of CCKG, whose validity 
has now largely been confirmed by subsequent studies (e.g., Gabuzda 2014b, 
2015a), suggested that the division between these regimes occurred at projected
distances of about 20~pc from the central AGN, and the results of our analysis
here bear this out.
This prediction will be tested more extensively by
future higher-resolution, long-wavelength radio observations that
will be made possible by the latest advances in radio telescopes
(ATA--256, EVLA, LOFAR, ASKAP, MeerKAT, SKA, VSOP--2; Gaensler et
al. 2004; Gaensler 2009; Perley et al. 2009; Hagiwara et al. 2009;
Law et al. 2011b). Further results from multi-wavelength
VLBA polarization observations of AGNs probing scales of
tens to hundreds of parsec may also shed more light on this question (e.g.,
Coughlan et al. 2010, Gabuzda et al. 2014a).

Finally, in a striking development, this picture has found
additional support from recent radio observations of the Galactic
center: Law, Brentjens, \& Novak (2011a) have reported a
powerful transverse CCW RM gradient ($\Delta$(RM)$\simeq
1100$~rad~m$^{-2}$ extending over 150~pc). Combining previous
observations of the area with their own observations, Law et al.
(2011a) find a "return" poloidal magnetic field in the central
$2^{\circ}$ of the Galactic center, with a dipolar configuration above and
below the Galactic plane. The measured RM values then flatten and
become positive in the central 30~pc, where the inner magnetic
field is expected to be superposed along the line of sight.
Additional observations by Pshirkov et al. (2011) have shown that
this CCW RM gradient extends to all Galactic latitudes above
and below the Galactic plane. This extended CCW RM gradient, the
flattening of the RM values towards the Galactic center, and the mapped 
dipolar magnetic field are all consistent with the expectations of the CB
mechanism. This supposes that the Galaxy had jets in the past,
which are now not detectable, but they have left a relic magnetic
field. In this case, the results of Pshirkov et al. (2011) amount to 
two more independent detections of CCW transverse RM gradients on large
scales (above and below the Galactic plane).
It is also interesting to note that this magnetic field structure is coincident
with the so called ``Fermi Bubbles"  (Su \& Finkbeiner 2012; Ackermann et al. 2014),
large-scale ($\sim$50$^o$) microwave and $\gamma$-ray structures above and below
the Galactic plane centered at the Galactic center.

In closing, it is clear that the data analyzed thus far show
intriguing trends that could be of cardinal importance to our
understanding of electromagnetic processes occurring in the jet--disk
systems of AGN. In particular, they support the global magnetic field
topology predicted by the Cosmic Battery model (Eqs.~(1)--(5)). If this
mechanism is indeed the dominant source of the initial magnetic fields in
AGN jets, this would indicate that AGNs could be an important source
of magnetic flux in the Universe.  The results
that we have presented here are undeniably based on small-number
statistics and require further confirmation from FR measurements
of additional radio sources. Our aim in reporting these results is to
bring these trends to the attention of the AGN community, in order
to spur further studies in this area, both observational and
theoretical.

\begin{acknowledgements}
We acknowledge insights and assistance with data provided by Drs.
Philip Best, Annalisa Bonafede, George Contopoulos, Federica Govoni, 
Christian Kaiser, and Preeti Kharb. We especially thank Annalisa Bonafede
and Federica Govoni for presenting us with the calibrated data for
A2142A and 5C4.152. We also acknowledge the assistance of
Antonios Nathanail in the preparation of Fig.~2. This work was
supported by the General Secretariat for Research and Technology
of Greece, the Irish Research Council (IRC) and the European Social 
Fund in the framework of Action
Excellence. 
We thank the referee whose comments have led to an expansion of the 
paper that helped improve the clarity and significance of our results.
\end{acknowledgements}

\newpage

\begin{table*}
\caption{Transverse RM gradients on decaparsec to kiloparsec scales}
\begin{tabular}{ccccccc}
\hline
No. &Object & $z$ & RM gradient & Projected distance & Instrument &References$^\dagger$\\
    & name  & & direction   & from core (pc) & and frequencies & \\
 (1) & (2) & (3) & (4) & (5)  & (6) & (7)\\ \hline
\multicolumn{7}{c}{Firm Gradients, Significances $\geq 3\sigma$}\\
\hline
1 & 0716+714  & 0.127 & CCW* & 3--35  & VLBA, 4.6-15~GHz  & {\bf Mahmud et al. (2013)} \\
  &           &       &      &        & 1.4-1.7~GHz  & Healy (2013) \\
2 & 0923+392  & 0.695 & CCW* & 20 & VLBA, 4.6-15~GHz  & {\bf Gabuzda et al. (2014b)} \\
3 & 5C4.114 (N)   & $> 0.023^a$ & CCW  & $> 2000$ &VLA, 1.4-4.9~GHz &  Bonafede et al. (2010);  \\
  &           &             &      &          &                 &  {\bf Gabuzda et al. (2015b)}  \\
4 & 5C4.114 (S)  & $> 0.023^a$ & CCW  & $> 1500$ &VLA, 1.4-4.9~GHz &  Bonafede et al. (2010);  \\
  &           &             &      &          &                 &  {\bf Gabuzda et al. (2015b)}  \\
5 & A2142A & 0.091 & CCW & to~$\simeq 10,000$ & VLA, 4.5-8.5~GHz &  Govoni et al. (2010); {\bf this paper}    \\
6 & 1652+398  & 0.034 & CCW  & 20 & VLBA, 8.4-1.7~GHz & {\bf Croke et al. (2010)} \\
7 & 1749+701  & 0.77  & CCW* & 75--100 & VLBA, 1.4-1.7~GHz & {\bf Mahmud et al. (2013)} \\
8 & 3C380     & 0.692 & CCW  & 70--210 & VLBA, 1.4-5.0~GHz & {\bf Gabuzda et al. (2014a)} \\
9 & 2037+511  & 1.687 & CCW* & 40 & VLBA, 4.6-15~GHz  & {\bf Gabuzda et al. (2014b)} \\
\hline
\multicolumn{7}{c}{Tentative Gradients}\\
\hline
1 & 0156$-$252 & 2.09 & CCW & 4000 & VLA, 1.4-8.5~GHz & Athreya et al. (1998) \\
2 & 3C120   & 0.033 & CCW & 25--80 & VLBA, 1.4-1.7~GHz & Coughlan et al. (2010); this paper \\
3 & M87  & 0.004 & CCW & 60, 960, 1400 & VLA, 8-43~GHz & Algaba et al. (2013)\\
4 & 5C4.152 & $\ldots^{a}$ & CCW & 15$^{\prime\prime}$  & VLA, 4.5-8.3~GHz &  Bonafede et al. (2010); this paper  \\
5 & Cen A & 0.0018 & CCW & 130,000 & ATCA, 1.3-1.5~MHz & Feain et al. (2009); this paper     \\
6 & 3C303 & 0.141 & CCW  & 20,000 & VLA, 1.4-8.5~GHz & Kronberg et al. (2011)   \\
7 & 3C465  & 0.0313 & CCW & 40,000--100,000 & VLA, 4.5-8.9~GHz &  Eilek \& Owen (2002)    \\
\hline
\hline
\multicolumn{6}{l}{$^\dagger$ References in bold indicate papers that present
quantitative analyses of the statistical significance of the}\\
\multicolumn{6}{l}{~~corresponding RM gradients.}\\
\multicolumn{6}{l}{*RM gradients in the CW direction are present closer
to the jet base.}\\
\multicolumn{6}{l}{$^a$~Exact redshift not available.}
\label{tab:gradients}
\end{tabular}
\end{table*}

\end{document}